\documentclass[11pt]{article}
\usepackage{fullpage,epsf,amssymb,amsthm,amsfonts,amsmath,latexsym,array,graphicx,extarrows,mathtools,mathstyle,mathrsfs,slashed,subcaption,amsbsy,csquotes,accents}
\usepackage[normalem]{ulem}
\usepackage{authblk}
\usepackage{cite}
\usepackage{color}

\usepackage{float}

\usepackage{hyperref}

\usepackage[export]{adjustbox}

\numberwithin{equation}{section}

\title{\bf{Euclidean thermal correlation functions in local QFT}} 

\author[1]{Peter Lowdon\thanks{lowdon@itp.uni-frankfurt.de}}

\affil[1]{Institut f\"{u}r Theoretische Physik, Johann Wolfgang Goethe-Universit\"{a}t, Max-von-Laue-Str. 1,  60438 Frankfurt am Main, Germany}

\date{}
\begin{document}

{\let\newpage\relax\maketitle}
\setcounter{page}{1}
\pagestyle{plain}

\begin{abstract}
\noindent
In this work we outline the general analytic characteristics satisfied by scalar correlation functions at finite temperature in local quantum field theory. We demonstrate that the locality of the fields in particular imposes significant constraints on the spectral structure of the theory, and that this enables the non-perturbative effects experienced by thermal particle states to be directly calculated from Euclidean correlation functions, avoiding the inverse problem.     
\end{abstract}

\newpage

\section{Introduction}
\label{intro}

Establishing the characteristics of quantum field theories (QFTs) in regimes of non-vanishing temperature is essential for describing the thermal effects experienced by particles as they move through a thermal medium, and this plays an important role in many physical systems such as the early Universe. In order to fully understand these thermal effects one ultimately requires a framework that does not depend on the specific coupling regime. Although relatively little is known about the general non-perturbative structure of QFTs at finite temperature, progress has been made for the simplest such class of theories, Hermitian scalar QFTs. In Refs.~\cite{Bros:1992ey,Bros:1995he,Bros:1998ua,Bros:1996mw,Bros:2001zs} important first steps were taken to demonstrate that a thermal QFT framework can be constructed using a series of physically motivated axioms, mirroring the approach of the already successful local QFT formulations at zero temperature, which have led to numerous foundational insights including the Spin-statistics Theorem, CPT Theorem, and the rigorous connection of Minkowski and Euclidean QFTs~\cite{Streater:1989vi,Haag:1992hx,Bogolyubov:1990kw}. A major difference with finite-temperature QFTs is that there no longer exists a unique ground state, but instead a thermal background state $|\Omega_{\beta}\rangle$ at temperature $T=1/\beta$. The appearance of a non-vanishing temperature defines a privileged reference frame and hence Lorentz symmetry is lost, which is reflected in the assumption that the fields $\phi(x)$ no longer transform covariantly under the full Poincar\'{e} group\footnote{Generally it is assumed that the fields transform under a unitary representation of some subgroup, which for the purposes of this work we will take to include spacetime translations and rotations.}. Nevertheless, there are several key characteristics that remain unchanged, including locality: $\left[\phi(x),\phi(y)\right]=0$ for $(x-y)^{2}<0$, and the fact that states are constructed by acting with the field operators on $|\Omega_{\beta}\rangle$. This implies that the thermal correlation functions $\langle \Omega_{\beta}|\phi(x_{1})\cdots\phi(x_{n})|\Omega_{\beta}\rangle$ encode all of the dynamical properties of the theory, and explains why they are central to understanding the characteristics of finite-temperature QFTs. \\  

\noindent
Of the non-perturbative techniques that exist to perform calculations at finite temperature, including lattice QFT, and functional methods such as the functional renormalisation group (FRG) and Dyson-Schwinger equations (DSEs), most are either restricted to, or optimised for, calculations in imaginary time. Therefore, in order to provide a correct physical interpretation of the data generated using these methods it is essential to understand the non-perturbative characteristics of Euclidean thermal correlation functions, and their connection to the physical Minkowski quantities. However, little is known about these characteristics, and this often means that challenging computational and theoretical obstacles must be overcome in order to determine meaningful observables. A particularly important example is the extraction of spectral information from Euclidean data, which necessarily requires the performance of a non-unique inversion, the so-called \textit{inverse problem}. In the context of lattice calculations of physical theories like quantum chromodynamics (QCD) this problem was initially studied in Ref.~\cite{Karsch:1986cq}. Several techniques have since been adapted and developed in order to circumvent this problem, including perhaps most prominently the Maximum Entropy Method~\cite{Jarrell:1996rrw,Asakawa:2000tr}, which was first applied to interpret lattice QCD data in Ref.~\cite{Nakahara:1999vy}. Although these techniques have helped advance the understanding of real-time spectral properties of finite-temperature QFTs~\cite{Jarrell:1996rrw,Asakawa:2000tr,Nakahara:1999vy,Meyer:2011gj,Haas:2013hpa,Tripolt:2018xeo}, the necessary assumptions required for their application leads to significant qualitative and quantitative uncertainties. The goal of this work is to help address this problem by using the full constraints of local QFT in order to establish the general spectral properties of scalar Euclidean thermal correlation functions. \\

\noindent
The remainder of this paper is structured as follows: in Sec.~\ref{gen_prop} we detail the non-perturbative constraints satisfied by two-point correlation functions of real scalar fields at finite temperature. In Sec.~\ref{euclidean} we determine how these constraints affect the characteristics in Euclidean spacetime, and in particular explore the consequences of massive particle states in the zero-temperature theory on the structure of the imaginary-time thermal correlation functions. Finally, in Sec.~\ref{concl} we summarise our main findings.

\section{Non-perturbative structure of thermal correlation functions}
\label{gen_prop}

In this section we outline the major constraints imposed on the structure of thermal correlation functions of real scalar fields due to the assumptions of local QFT, in particular locality, translational invariance, and thermal equilibrium. For the purposes of this work we will focus on the thermal correlation functions involving two field operators.

\subsection{Thermal two-point function}
\label{two-point}

In the case of the thermal two-point function, translational invariance implies that the correlation function depends only on the difference variable $x-y$, and hence one can define $ \langle \Omega_{\beta}| \phi(x)\phi(y) |\Omega_{\beta} \rangle=\mathcal{W}_{\beta}(x-y)$. The condition of thermal equilibrium is expressed via the Kubo-Martin-Schwinger (KMS) condition~\cite{Haag:1967sg}
\begin{align}
&\langle \Omega_{\beta}|\phi(x)\phi(y)|\Omega_{\beta}\rangle   = \langle \Omega_{\beta}|\phi(y) \phi(x+i(\beta,\vec{0}))|\Omega_{\beta}\rangle,
\label{KMS}
\end{align}
which in light of the translational invariance implies that in momentum space
\begin{align}
\widetilde{\mathcal{W}}_{\beta}(p) = e^{\beta p_{0}}\widetilde{\mathcal{W}}_{\beta}(-p).
\label{KMS_p}
\end{align}
In the case of real scalar fields, Eq.~\eqref{KMS_p} can be rewritten in terms of the thermal commutator $C_{\beta}(x-y) =\langle \Omega_{\beta}| \left[\phi(x),\phi(y)\right]|\Omega_{\beta} \rangle$, resulting in the condition
\begin{align}
\widetilde{\mathcal{W}}_{\beta}(p) = \frac{\widetilde{C}_{\beta}(p)}{1-e^{-\beta p_{0}}}.
\label{W_rep}
\end{align} 
The characteristics outlined so far are well-known features of finite-temperature QFT that have been instrumental in forming the conventional treatment of this subject~\cite{Kapusta:2006pm,Bellac:2011kqa}. Although it has long been understood that the thermal commutator $\widetilde{C}_{\beta}(p)$, or \textit{spectral function}, is an important quantity for understanding the thermal effects experienced by particle states at finite temperature, the significant constraints imposed on $\widetilde{C}_{\beta}(p)$ due to the locality of the fields has largely been overlooked. In Ref.~\cite{Bros:1992ey} it was first demonstrated that locality implies that $\widetilde{C}_{\beta}(p)$ can be written in the general form
\begin{align}
\widetilde{C}_{\beta}(p_{0},\vec{p}) = \int_{0}^{\infty} \! ds \int \! \frac{d^{3}\vec{u}}{(2\pi)^{2}} \ \epsilon(p_{0}) \, \delta\!\left(p^{2}_{0} - (\vec{p}-\vec{u})^{2} - s \right) \widetilde{D}_{\beta}(\vec{u},s),
\label{spec_rep}
\end{align}
which upon combining with Eq.~\eqref{W_rep} leads to the following spectral representation for the thermal two-point function 
\begin{align}
\widetilde{\mathcal{W}}_{\beta}(p_{0},\vec{p})  = \int_{0}^{\infty} \! ds \int \! \frac{d^{3}\vec{u}}{(2\pi)^{2}} \  \frac{\epsilon(p_{0})  \delta\!\left(p^{2}_{0} - (\vec{p}-\vec{u})^{2} - s \right)}{1-e^{-\beta p_{0}}} \, \widetilde{D}_{\beta}(\vec{u},s).
\label{W_spec_rep}
\end{align}
A remarkable characteristic of Eq.~\eqref{W_spec_rep} is that the dynamical effects of the thermal background medium are \textit{entirely} controlled by $\widetilde{D}_{\beta}(\vec{u},s)$. The quantity $\widetilde{D}_{\beta}(\vec{u},s)$ has the interpretation of a thermal spectral density, since in the limit of vanishing temperature ($\beta\rightarrow \infty$) the Poincar\'{e} covariance of the fields is restored, $|\Omega_{\beta} \rangle$ approaches the vacuum state $|0\rangle$, and this implies
\begin{align}
\widetilde{D}_{\beta}(\vec{u},s) \  \xlongrightarrow{\beta\rightarrow \infty}{} \ (2\pi)^{3} \delta^{3}(\vec{u})\, \rho(s),
\label{D_lim}
\end{align}
where $\rho(s)$ is the spectral density of the zero-temperature theory. In the $\beta\rightarrow \infty$ limit the thermal two-point function therefore approaches the standard K\"{a}ll\'{e}n-Lehmann spectral representation of the vacuum theory~\cite{Kallen:1952zz,Lehmann:1954xi} 
\begin{align}
\widetilde{\mathcal{W}}_{\beta}(p_{0},\vec{p}) \  \xlongrightarrow{\beta\rightarrow \infty}{} \ 2\pi \, \theta(p_{0}) \! \int_{0}^{\infty} \! ds \ \delta\!\left(p^{2} - s \right)\rho(s),
\label{KL_rep}
\end{align}
and hence Eq.~\eqref{W_spec_rep} corresponds to the $T>0$ generalisation of this representation.

\subsection{Retarded and advanced thermal propagators}
\label{ret_adv}

The retarded $r_{\beta}(x) = i\theta(x^{0})C_{\beta}(x)$ and advanced $a_{\beta}(x) = -i\theta(-x^{0})C_{\beta}(x)$ thermal propagators also play an important role in finite-temperature QFTs. In momentum space, the retarded propagator has the integral form
\begin{align}
\widetilde{r}_{\beta}(p_{0},\vec{p}) = i\int_{-\infty}^{\infty} \frac{dq_{0}}{2\pi} \frac{i}{(p_{0}-q_{0})+i\epsilon}\widetilde{C}_{\beta}(q_{0},\vec{p}),
\label{ret_int}
\end{align}
and admits an analytic continuation $\widetilde{R}_{\beta}(k_{0},\vec{p})$ to complex energy values $k_{0} \in \mathbb{C}$, with the real-energy propagator recovered in the limit $\lim_{\epsilon\rightarrow 0^{+}}\widetilde{R}_{\beta}(p_{0}+i\epsilon,\vec{p})$. An equivalent structure also exists for the advanced propagator, where instead $\widetilde{a}_{\beta}(p_{0},\vec{p})=\lim_{\epsilon\rightarrow 0^{+}}\widetilde{A}_{\beta}(p_{0}-i\epsilon,\vec{p})$, with $\widetilde{A}_{\beta}(k_{0},\vec{p})$ the analytic continuation. In general, the analytic functions $\widetilde{R}_{\beta}(k)$ and $\widetilde{A}_{\beta}(k)$ are different, but if $\widetilde{C}_{\beta}(p)$ vanishes in some region of energy-momentum space it follows that they can in fact be analytically continued into one another~\cite{Bros:2001zs}. Hence in this case there exists a \textit{single} propagator function $\widetilde{G}_{\beta}(k_{0},\vec{p})$ for which $\widetilde{r}_{\beta}(p)$ and $\widetilde{a}_{\beta}(p)$ are recovered as boundary values. Due to Eqs.~\eqref{spec_rep} and~\eqref{ret_int} one can therefore demonstrate that this propagator has the spectral representation   
\begin{align}
\widetilde{G}_{\beta}(k_{0},\vec{p}) = -\int_{0}^{\infty} \! ds \int \frac{d^{3}\vec{u}}{(2\pi)^{3}}   \, \frac{\widetilde{D}_{\beta}(\vec{u},s)}{k^{2}_{0} - (\vec{p}-\vec{u})^{2} - s }.
\label{spec_rep_RA_prop_1}
\end{align}
It turns out that the inverse spatial Fourier transform of $\widetilde{G}_{\beta}(k_{0},\vec{p})$ is a particularly important quantity for highlighting the structural properties of the thermal two-point function in Euclidean spacetime, as will be discussed in Sec~\ref{euclidean}. Using Eq.~\eqref{spec_rep_RA_prop_1}, one finds that
\begin{align}
\undertilde{G}_{\beta}(k_{0},\vec{x}) = \int \frac{d^{3}\vec{p}}{(2\pi)^{3}} e^{i \vec{p}\cdot \vec{x}}\, \widetilde{G}_{\beta}(k_{0},\vec{p}) = \frac{1}{4\pi |\vec{x}|}\int_{0}^{\infty} ds \, e^{-|\vec{x}|\sqrt{s-k_{0}^{2}}} D_{\beta}(\vec{x},s), 
\label{D_rel}
\end{align}
where $D_{\beta}(\vec{x},s)$ is the position space thermal spectral density, and $k_{0}$ is restricted such that either $k_{0}^{2} \not\in \mathbb{R}$, or $\text{Re}\!\left(k_{0}^{2}\right) \leq s$.

\subsection{Spectral decomposition}
\label{spec_decomp}

As outlined in Secs.~\ref{two-point} and~\ref{ret_adv}, by virtue of Eq.~\eqref{W_spec_rep} the thermal spectral density is key to understanding the structure of thermal correlation functions, and hence the dynamical properties of QFTs at finite temperature. Although there remain open questions regarding the precise properties of $\widetilde{D}_{\beta}(\vec{u},s)$, a reasonable assumption is that the singular structure in the variable $s$ is preserved relative to the vacuum theory, which implies that discrete and continuous contributions can be similarly decomposed. Therefore, if a theory contains a single particle state of mass $m$ at $T=0$, one can write $\widetilde{D}_{\beta}(\vec{u},s)$ in the following manner~\cite{Bros:2001zs}
\begin{align}
\widetilde{D}_{\beta}(\vec{u},s)= \widetilde{D}_{m,\beta}(\vec{u})\, \delta(s-m^{2}) + \widetilde{D}_{c, \beta}(\vec{u},s),
\label{decomp}
\end{align} 
where $\widetilde{D}_{c, \beta}(\vec{u},s)$ is continuous in the variable $s$. Besides the mathematical considerations, Eq.~\eqref{decomp} also provides a natural physical description of particle states moving within a thermal medium. For $T>0$ the function $\widetilde{D}_{m,\beta}(\vec{u})$ is non-trivial, and due to the convolution structure of Eq.~\eqref{W_spec_rep} this causes the thermal two-point function to have contributions outside of the mass shell $p^{2}=m^{2}$. The zero-temperature mass of the particle $m$ is therefore screened by thermal effects characterised by the damping factor $\widetilde{D}_{m,\beta}(\vec{u})$, the behaviour of which is determined by the underlying dynamics between the particle and the constituents in the background state. The properties of damping factors in specific models were first explored in Ref.~\cite{Bros:2001zs}, and more recently in Ref.~\cite{Lowdon:2021ehf}, where it was demonstrated that these quantities are important in the calculation of observables such as the shear viscosity. In Sec.~\eqref{euclidean} we will explore the effect that these various non-perturbative real-time characteristics have on the structure of the thermal correlation functions in Euclidean spacetime.

\section{Thermal correlation functions in Euclidean spacetime}
\label{euclidean}

\subsection{General analytic properties}
\label{gen_analytic}

As outlined in Sec.~\ref{gen_prop}, the absence of full Lorentz invariance has significant implications for the properties of thermal correlation functions. It is therefore not surprising that the analytic continuation to imaginary time $\tau$ is a more subtle issue than in the vacuum case. It is well known that by virtue of the KMS condition and locality the corresponding Euclidean two-point function $\mathcal{W}_{E}(\tau, \vec{x})$ must be $\beta$-periodic. As such, $\mathcal{W}_{E}$ can be represented in terms of a Fourier series expansion
\begin{align}
\mathcal{W}_{E}(\tau, \vec{x}) = \frac{1}{\beta} \! \sum_{N=-\infty}^{\infty} \! w_{N}(\vec{x}) \, e^{\frac{2\pi i N}{\beta}\tau},
\label{Fourier_series}
\end{align} 
with Fourier coefficients $w_{N}(\vec{x})$. Furthermore, due to the analytic properties of $\mathcal{W}_{E}$ it follows that the Fourier transform of these coefficients are related to the analytic continuation of the real-time retarded propagator as follows~\cite{Bros:1996mw}:
\begin{align}
\widetilde{w}_{N}(\vec{p}) = \widetilde{R}_{\beta}(i\omega_{N},\vec{p}), \hspace{5mm}  N \geq 0,
\label{ret_1}
\end{align} 
where $\omega_{N}= \frac{2\pi N}{\beta}$. An analogous relation also holds for $N \leq 0$, but instead involving the analytic continuation of the advanced propagator $\widetilde{A}_{\beta}(k)$. As detailed in Sec.~\ref{ret_adv}, if the retarded and advanced propagators coincide in some region then they form the components of a single analytic propagator function $\widetilde{G}_{\beta}(k)$. In this case Eq.~\eqref{ret_1} generalises to 
\begin{align}
\widetilde{w}_{N}(\vec{p}) = \widetilde{G}_{\beta}(i\omega_{N},\vec{p}), \hspace{5mm}  N \in \mathbb{Z}.
\label{Fourer_coeff_rel}
\end{align}
These analytic relations emphasise the deep connection between the imaginary and real-time formulations of QFT at finite temperature. In particular, given knowledge of the propagator at the Matsubara frequencies $k_{0}= i\omega_{N}$, one can fully reconstruct the Euclidean two-point function $\mathcal{W}_{E}$.

\subsection{Connection to the thermal spectral density}
\label{therm_conn}
   
Given the analytic relations outlined in Sec.~\ref{gen_analytic} one can now use the spectral representations of the propagators in Sec.~\ref{ret_adv} in order to explore the relationship between the thermal spectral density and the Euclidean two-point function. In particular, combining Eqs.~\eqref{D_rel} and~\eqref{Fourer_coeff_rel} gives
\begin{align}
w_{N}(\vec{x}) =\frac{1}{4\pi |\vec{x}|}\int_{0}^{\infty} \! ds \ e^{-|\vec{x}|\sqrt{s+\omega_{N}^{2}}} D_{\beta}(\vec{x},s), 
\label{fourier_coeff_matsu}
\end{align}
where $N \in \mathbb{Z}$. Equation~\eqref{fourier_coeff_matsu} explicitly demonstrates that the Fourier coefficients, and hence the full Euclidean two-point function, can be directly derived from the form of $D_{\beta}(\vec{x},s)$. \\

\noindent
Just like at zero temperature, the specific properties of the thermal spectral density depend on the dynamics of the theory in question. However, as outlined in Sec.~\ref{spec_decomp}, in Ref.~\cite{Bros:2001zs} it was proposed that if the zero-temperature theory contains a single massive particle state in the spectrum, then $D_{\beta}(\vec{x},s)$ should satisfy the decomposition in Eq.~\eqref{decomp}. In this case, it follows from Eq.~\eqref{fourier_coeff_matsu} that the $N=0$ Fourier coefficient has the form
\begin{align}
w_{0}(\vec{x}) =   \frac{1}{4\pi |\vec{x}|}\left[ D_{m,\beta}(\vec{x}) e^{-|\vec{x}|m}  +\int_{0}^{\infty} \! ds \, e^{-|\vec{x}|\sqrt{s}} D_{c,\beta}(\vec{x},s) \right],  
\label{fourier_coeff_ansatz}
\end{align}
where $m$ is the zero-temperature mass of the particle state. An immediate consequence of Eq.~\eqref{fourier_coeff_ansatz} is that the $s$-dependence of $D_{c,\beta}(\vec{x},s)$ plays an essential role in determining the relative contributions of the discrete and continuous spectral contributions to $w_{0}(\vec{x})$. In the specific case that $D_{c,\beta}(\vec{x},s)$ is non-vanishing in the region $s \geq s_{c}>m^{2}$, and the gap $s_{c}-m^{2}$ is sufficiently large, this implies:
\begin{align}
&\textit{The damping factor $D_{m,\beta}(\vec{x})$ will dominate the behaviour of $w_{0}(\vec{x})$, and this} \nonumber \\
&\quad\quad\quad \textit{domination will be especially pronounced for larger values of $|\vec{x}|$.} 
\label{dom_cond}  
\end{align}
This condition arises from the fact that $D_{c,\beta}(\vec{x},s)$ can only grow at most polynomially in $|\vec{x}|$ since it is a tempered distribution\footnote{That $D_{c,\beta}(\vec{x},s)$ is a distribution follows from the fundamental local QFT assumption that fields $\phi(x)$ are operator-valued tempered distributions~\cite{Streater:1989vi,Haag:1992hx}.} in the variable $\vec{x}$, and therefore regardless of the specific $\vec{x}$-behaviour the second term in Eq.~\eqref{fourier_coeff_ansatz} will be increasingly suppressed relative to the first as the gap $s_{c}-m^{2}$ becomes larger. If the continuous component \textit{also} possesses a factorised form $D_{c,\beta}(\vec{x},s)=\mathcal{D}_{c,\beta}(\vec{x})\mathcal{C}(s)$, then its relative contribution to $w_{0}(\vec{x})$ will be dictated by the $T=0$ spectral properties of the theory, since in this case the gap $s_{c}-m^{2}$ is temperature independent. \\

\noindent
The dominance of the discrete contribution is reminiscent of what one finds in the zero-temperature QCD sum rule technique, where the application of a Borel transformation to hadronic amplitudes leads to the exponential suppression of the continuum states~\cite{Shifman:1978by}. The fundamental difference here though is that the suppression effect in Eq.~\eqref{fourier_coeff_ansatz} occurs by virtue of the Euclidean analytic structure alone\footnote{In the Minkowski spacetime case it has been shown that $D_{m,\beta}(\vec{x})$ instead dominates the real-time two-point function $\mathcal{W}_{\beta}(x_{0}, \vec{x})$ in the \textit{asymptotic} region $x_{0}\rightarrow \pm\infty$~\cite{Bros:2001zs}.}. This is significant because it demonstrates that Euclidean spacetime information provides direct insights into the in-medium effects experienced by thermal particle states. In particular, combining Eq.~\eqref{fourier_coeff_ansatz} with the conclusion in Eq.~\eqref{dom_cond} it follows that for a theory with a single massive particle state at $T=0$, and a large gap $s_{c}-m^{2}$, the corresponding damping factor of this state can be directly estimated from the $N=0$ Fourier coefficient  
\begin{align}
D_{m,\beta}(\vec{x}) \sim 4\pi |\vec{x}|\, e^{|\vec{x}|m} w_{0}(\vec{x}).
\label{Fourier_coeff_damping_rel}
\end{align} 
In principle one could also attempt to construct analogous relations to Eq.~\eqref{Fourier_coeff_damping_rel} for $N>0$. However, in this case the dominance of the damping factor contribution to Eq.~\eqref{fourier_coeff_matsu} is no longer guaranteed, since the suppression of $D_{c, \beta}(\vec{x},s)$ is increasingly diminished for larger values of $N$, and for higher temperatures. In this sense, the $N=0$ Fourier coefficient $w_{0}(\vec{x})$ is optimal for extracting the behaviour of the damping factor. In Sec.~\ref{extract} we will demonstrate how this result can be used to extract the form of the damping factor given Euclidean correlation function data in either position or momentum space. 
 
\subsection{Extracting thermal spectral densities from Euclidean data}
\label{extract}

In light of the analysis in Sec.~\ref{therm_conn}, if a local finite-temperature QFT possesses a scalar particle state with mass $m$ at $T=0$, and the gap with the continuum is sufficiently large, then the associated damping factor of this state $D_{m,\beta}(\vec{x})$ can be estimated from the Euclidean scalar two-point function $\mathcal{W}_{E}(\tau, \vec{x})$. In particular, combining Eqs.~\eqref{Fourier_series} and~\eqref{Fourier_coeff_damping_rel} it follows that
\begin{align}
D_{m,\beta}(\vec{x}) \sim  4\pi |\vec{x}| \, e^{|\vec{x}|m} \! \int_{-\frac{\beta}{2}}^{\frac{\beta}{2}} d\tau \, \mathcal{W}_{E}(\tau, \vec{x}).
\label{Fourier_coeff_dampingx}
\end{align}
Upon taking the Fourier transform of $D_{m,\beta}(\vec{x})$, the full contribution of the scalar particle state to the spectral function $\widetilde{C}_{\beta}(p_{0},\vec{p})$, and hence $\rho(\omega)=\widetilde{C}_{\beta}(\omega,\vec{p}=0)$, can then be directly computed by setting $\widetilde{D}_{\beta}(\vec{u},s)\rightarrow \widetilde{D}_{m,\beta}(\vec{u})\, \delta(s-m^{2})$ in Eq.~\eqref{spec_rep}. In the standard approach, the zero-momentum spectral function $\rho(\omega)$ is determined from the \textit{spatially}-integrated Euclidean two-point function $\mathcal{W}_{E}(\tau)=\int d^{3}x \, \mathcal{W}_{E}(\tau, \vec{x})$ via the relation  
\begin{align}
\mathcal{W}_{E}(\tau) = \int_{0}^{\infty}  \! \frac{d\omega}{2\pi} \,  \frac{\cosh\left[\left(\tfrac{\beta}{2}-|\tau| \right)\omega\right] }{\sinh\left(\tfrac{\beta}{2}\omega\right)} \, \rho(\omega).
\label{standard_rel}
\end{align}
A fundamental characteristic of this approach is that one has to deal with the fact that $\rho(\omega)$ cannot be uniquely reconstructed from Eq.~\eqref{standard_rel} given knowledge of $\mathcal{W}_{E}(\tau)$, which is the realisation of the inverse problem outlined in Sec.~\ref{intro}. By contrast, the direct computation of particle contributions to $\rho(\omega)$ via Eq.~\eqref{Fourier_coeff_dampingx} avoids this problem. \\

\noindent 
So far we have only considered the situation where one has Euclidean data in position space. In non-perturbative approaches such as the FRG and DSEs one instead generally calculates the form of correlation functions in momentum space, in particular the analytic propagator $\widetilde{G}_{\beta}(k_{0},\vec{p})$. In this case, one can directly apply Eq.~\eqref{Fourer_coeff_rel}. After setting $N=0$, and taking the inverse Fourier transform, it then follows from Eq.~\eqref{Fourier_coeff_damping_rel} that\footnote{Here we use the fact that the propagator depends only on the absolute value of $\vec{p}$, which follows from the assumption of rotational invariance.}
\begin{align}
D_{m,\beta}(\vec{x}) \sim    e^{|\vec{x}|m}\int_{0}^{\infty} \! \frac{d|\vec{p}|}{2\pi} \ 4|\vec{p}| \sin\!\left(|\vec{p}||\vec{x}|\right) \, \widetilde{G}_{\beta}(0,|\vec{p}|).
\label{Fourier_coeff_dampingp}
\end{align} 
As with Eq.~\eqref{Fourier_coeff_dampingx}, one can then use this expression to directly calculate particle contributions to the full spectral function, and hence $\rho(\omega)$. Taken together, Eqs.~\eqref{Fourier_coeff_dampingx} and~\eqref{Fourier_coeff_dampingp} provide a new approach with which one can both calculate and understand the non-perturbative effects experienced by particle states in a thermal medium. The applications of both of these relations are the subject forthcoming papers. In particular, in Ref.~\cite{Lowdon:2022ird} Eq.~\eqref{Fourier_coeff_dampingp} is applied in order to determine the explicit functional form and temperature dependence of the damping factor of pion states from FRG-generated data in the quark-meson model. With the analytic results derived in Ref.~\cite{Lowdon:2021ehf} this expression is then used to calculate non-perturbative observables, including the shear viscosity arising from the pion states.

\section{Conclusions} 
\label{concl}

In this work we outline how the non-perturbative characteristics of local QFT at finite temperature manifest themselves in Euclidean spacetime for real scalar fields, focussing in particular on the situation in which there exists massive particle states in the zero-temperature limit. We demonstrate for the first time that the thermal generalisation of the K\"{a}ll\'{e}n-Lehmann representation in the real-time theory has a significant impact on the structure of the Euclidean two-point function, and that ultimately this enables the non-perturbative effects experienced by thermal particle states to be directly calculated from Euclidean data, avoiding the well-known inverse problem. Although this study concerns the simplest class of finite-temperature QFTs, these results can in principle be generalised to theories with fields of higher spin, or to systems with a background state with non-vanishing density. These generalisations will be the subject of future investigations. The results in this work therefore represent a key step towards the characterisation of non-perturbative in-medium effects in more complex physical theories.

\section*{Acknowledgements}
The author would like to thank Jan Pawlowski and Arno Tripolt for useful discussions and input. The work of P.~L. is supported by the Deutsche Forschungsgemeinschaft (DFG, German Research Foundation) through the Collaborative Research Center CRC-TR 211 ``Strong-interaction matter under extreme conditions'' -- Project No. 315477589-TRR 211.

\bibliographystyle{JHEP}

\bibliography{refs}

\providecommand{\href}[2]{#2}\begingroup\raggedright\begin{thebibliography}{10}

\bibitem{Bros:1992ey}
J.~Bros and D.~Buchholz, \emph{{Particles and propagators in relativistic
  thermo field theory}}, \href{https://doi.org/10.1007/BF01565114}{\emph{Z.
  Phys. C} {\bfseries 55} (1992) 509}.

\bibitem{Bros:1995he}
J.~Bros and D.~Buchholz, \emph{{Relativistic KMS condition and Kallen-Lehmann
  type representations of thermal propagators}},  in \emph{{4th Workshop on
  Thermal Field Theories and Their Applications}}, pp.~103--110, 8, 1995,
  \href{https://arxiv.org/abs/hep-th/9511022}{{\ttfamily hep-th/9511022}}.

\bibitem{Bros:1998ua}
J.~Bros and D.~Buchholz, \emph{{Towards a relativistic KMS condition}},
  \href{https://doi.org/10.1016/0550-3213(94)00298-3}{\emph{Nucl. Phys. B}
  {\bfseries 429} (1994) 291}
  [\href{https://arxiv.org/abs/hep-th/9807099}{{\ttfamily hep-th/9807099}}].

\bibitem{Bros:1996mw}
J.~Bros and D.~Buchholz, \emph{{Axiomatic analyticity properties and
  representations of particles in thermal quantum field theory}}, {\emph{Ann.
  Inst. H. Poincare Phys. Theor.} {\bfseries 64} (1996) 495}
  [\href{https://arxiv.org/abs/hep-th/9606046}{{\ttfamily hep-th/9606046}}].

\bibitem{Bros:2001zs}
J.~Bros and D.~Buchholz, \emph{{Asymptotic dynamics of thermal quantum
  fields}}, \href{https://doi.org/10.1016/S0550-3213(02)00059-7}{\emph{Nucl.
  Phys. B} {\bfseries 627} (2002) 289}
  [\href{https://arxiv.org/abs/hep-ph/0109136}{{\ttfamily hep-ph/0109136}}].

\bibitem{Streater:1989vi}
R.~F. Streater and A.~S. Wightman, \emph{{PCT, spin and statistics, and all
  that}}. Redwood City, USA: Addison-Wesley, 1989.

\bibitem{Haag:1992hx}
R.~Haag, \emph{{Local quantum physics: Fields, particles, algebras}}. Berlin,
  Germany: Springer, 1992.

\bibitem{Bogolyubov:1990kw}
N.~N. Bogolyubov, A.~A. Logunov, A.~I. Oksak and I.~T. Todorov, \emph{{General
  Principles of Quantum Field Theory}}. Dordrecht, Netherlands: Kluwer, 1990.

\bibitem{Karsch:1986cq}
F.~Karsch and H.~W. Wyld, \emph{{Thermal Green's Functions and Transport
  Coefficients on the Lattice}},
  \href{https://doi.org/10.1103/PhysRevD.35.2518}{\emph{Phys. Rev. D}
  {\bfseries 35} (1987) 2518}.

\bibitem{Jarrell:1996rrw}
M.~Jarrell and J.~E. Gubernatis, \emph{{Bayesian inference and the analytic
  continuation of imaginary-time quantum Monte Carlo data}},
  \href{https://doi.org/10.1016/0370-1573(95)00074-7}{\emph{Phys. Rept.}
  {\bfseries 269} (1996) 133}.

\bibitem{Asakawa:2000tr}
M.~Asakawa, T.~Hatsuda and Y.~Nakahara, \emph{{Maximum entropy analysis of the
  spectral functions in lattice QCD}},
  \href{https://doi.org/10.1016/S0146-6410(01)00150-8}{\emph{Prog. Part. Nucl.
  Phys.} {\bfseries 46} (2001) 459}
  [\href{https://arxiv.org/abs/hep-lat/0011040}{{\ttfamily hep-lat/0011040}}].

\bibitem{Nakahara:1999vy}
Y.~Nakahara, M.~Asakawa and T.~Hatsuda, \emph{{Hadronic spectral functions in
  lattice QCD}}, \href{https://doi.org/10.1103/PhysRevD.60.091503}{\emph{Phys.
  Rev. D} {\bfseries 60} (1999) 091503(R)}
  [\href{https://arxiv.org/abs/hep-lat/9905034}{{\ttfamily hep-lat/9905034}}].

\bibitem{Meyer:2011gj}
H.~B. Meyer, \emph{{Transport Properties of the Quark-Gluon Plasma: A Lattice
  QCD Perspective}},
  \href{https://doi.org/10.1140/epja/i2011-11086-3}{\emph{Eur. Phys. J. A}
  {\bfseries 47} (2011) 86} [\href{https://arxiv.org/abs/1104.3708}{{\ttfamily
  1104.3708}}].

\bibitem{Haas:2013hpa}
M.~Haas, L.~Fister and J.~M. Pawlowski, \emph{{Gluon spectral functions and
  transport coefficients in Yang--Mills theory}},
  \href{https://doi.org/10.1103/PhysRevD.90.091501}{\emph{Phys. Rev. D}
  {\bfseries 90} (2014) 091501(R)}
  [\href{https://arxiv.org/abs/1308.4960}{{\ttfamily 1308.4960}}].

\bibitem{Tripolt:2018xeo}
R.-A. Tripolt, P.~Gubler, M.~Ulybyshev and L.~Von~Smekal, \emph{{Numerical
  analytic continuation of Euclidean data}},
  \href{https://doi.org/10.1016/j.cpc.2018.11.012}{\emph{Comput. Phys. Commun.}
  {\bfseries 237} (2019) 129}
  [\href{https://arxiv.org/abs/1801.10348}{{\ttfamily 1801.10348}}].

\bibitem{Haag:1967sg}
R.~Haag, N.~M. Hugenholtz and M.~Winnink, \emph{{On the Equilibrium states in
  quantum statistical mechanics}},
  \href{https://doi.org/10.1007/BF01646342}{\emph{Commun. Math. Phys.}
  {\bfseries 5} (1967) 215}.

\bibitem{Kapusta:2006pm}
J.~I. Kapusta and C.~Gale, \emph{{Finite-temperature Field Theory: Principles
  and applications}}, Cambridge Monographs on Mathematical Physics. Cambridge
  University Press, 2011.

\bibitem{Bellac:2011kqa}
M.~L. Bellac, \emph{{Thermal Field Theory}}, Cambridge Monographs on
  Mathematical Physics. Cambridge University Press, 3, 2011.

\bibitem{Kallen:1952zz}
G.~{K\"{a}ll\'{e}n}, \emph{{On the definition of the Renormalization Constants
  in Quantum Electrodynamics}}, {\emph{Helv. Phys. Acta} {\bfseries 25} (1952)
  417}.

\bibitem{Lehmann:1954xi}
H.~Lehmann, \emph{{On the Properties of propagation functions and
  renormalization contants of quantized fields}},
  \href{https://doi.org/10.1007/BF02783624}{\emph{Nuovo Cim.} {\bfseries 11}
  (1954) 342}.

\bibitem{Lowdon:2021ehf}
P.~Lowdon, R.-A. Tripolt, J.~M. Pawlowski and D.~H. Rischke, \emph{{Spectral
  representation of the shear viscosity for local scalar QFTs at finite
  temperature}}, \href{https://doi.org/10.1103/PhysRevD.104.065010}{\emph{Phys.
  Rev. D} {\bfseries 104} (2021) 065010}
  [\href{https://arxiv.org/abs/2104.13413}{{\ttfamily 2104.13413}}].

\bibitem{Shifman:1978by}
M.~A. Shifman, A.~I. Vainshtein and V.~I. Zakharov, \emph{{QCD and Resonance
  Physics: Applications}},
  \href{https://doi.org/10.1016/0550-3213(79)90023-3}{\emph{Nucl. Phys. B}
  {\bfseries 147} (1979) 448}.

\bibitem{Lowdon:2022ird}
P.~Lowdon and R.-A. Tripolt, \emph{{Real-time observables from Euclidean
  thermal correlation functions}},
  \href{https://doi.org/10.1103/PhysRevD.106.056006}{\emph{Phys. Rev. D}
  {\bfseries 106} (2022) 056006}
  [\href{https://arxiv.org/abs/2202.09142}{{\ttfamily 2202.09142}}].

\end{thebibliography}\endgroup

\end{document}